\documentclass[fleqn,usenatbib]{mnras}

\usepackage{newtxtext,newtxmath}

\usepackage[T1]{fontenc}
\usepackage{ae,aecompl}

\usepackage{graphicx}	
\usepackage{amsmath}	
\usepackage{amssymb}	
\usepackage{times,varioref,multirow,textcomp,comment,multicol}
\usepackage{xspace,float}
\usepackage[usenames,dvipsnames,svgnames,hyperref]{xcolor}
\usepackage[percent]{overpic}
\usepackage{empheq,bm}
\usepackage{varwidth}

\title[Dark Matter Annihilation Feedback II]{Dark Matter Annihilation Feedback in Cosmological Simulations II: The Influence on Gas and Halo Structure}

\author[N. Iwanus et al.]{
N. Iwanus$^{1}$\thanks{E-mail: nikolas.iwanus@sydney.edu.au},
P. J. Elahi$^{2,3}$,
F. List$^{1}$
and G. F. Lewis$^{1}$
\\
$^{1}$Sydney Institute for Astronomy, School of Physics, A28, The University of Sydney, NSW 2006, Australia\\
$^{2}$International Centre for Radio Astronomy Research, University of Western Australia, 35 Stirling Highway, Crawley, WA 6009, Australia\\
$^{3}$ARC Centre of Excellence for All Sky Astrophysics in 3 Dimensions (ASTRO 3D)
}

\date{Accepted XXX. Received YYY; in original form ZZZ}

\pubyear{2018}
\begin{document}
\label{firstpage}
\pagerange{\pageref{firstpage}--\pageref{lastpage}}
\maketitle

\begin{abstract}
We present new cosmological hydrodynamic simulations that incorporate Dark Matter Annihilation Feedback (DMAF), whereby energy released from the annihilation of dark matter particles through decay channels such as photon or positron-electron pairs provide additional heating sources for local baryonic material.
For annihilation rates comparable to WIMP-like particles, we find that the key influence of DMAF is to inhibit gas accretion onto halos.  
Such diminished gas accretion early in the lifetimes of halos results in reduced gas fractions in smaller halos, and the delayed halo formation times of larger structures, suggesting that DMAF could impact the stellar age distribution in galaxies, and morphology of dwarfs. 
For a dark matter particle mass of $m_\chi\sim10$~MeV, there is a `critical halo mass' of $\sim10^{13}$ M$_{\odot}$ at $z=0$, below which there are large differences when compared to $\Lambda$CDM, such as a reduction in the abundance of halo structures as large as 25 percent, reduced gas content by 50 percent and central gas densities reduced down to 10 percent within halos of mass $\sim10^{12}$ M$_{\odot}$ but with increasing effects in smaller halos. 
Higher dark matter particle mass models have a smaller `critical halo mass'. For a $m_\chi\sim100$~MeV model, we find differences start appearing below halo masses of $\sim10^{12}$ M$_\odot$ and a $m_\chi\gtrsim 1$~GeV model, this mass scale lies below the resolution of our simulations, though we still observe changes in the  morphology of dwarf galaxies.
\end{abstract}




\section{Introduction}
The standard cosmological model, where the Universe is dominated by dark energy and cold dark matter, has been shown to provide excellent agreement with observations of the larger-scale cosmos
\citep[e.g.][]{Planck2016}. 
However, predictions from high-resolution numerical simulations based on $\Lambda$CDM cosmologies, appear to be in tension with the smaller-scale structure of the Universe \citep{Bullock2017Review}. 
For example, numerical dark matter halos possess sharp central cusps in their density profiles, while rotation curves of nearby dwarf galaxies indicate that they reside in cored halos \citep{Kravtsov1998,Oh2015}. 
Furthermore, numerical universes appear to overproduce satellites around Milky Way-like galaxies \citep{Klypin1999,Moore1999}, an issue that could be solved with baryon feedback effects like UV radiation \citep{Thoul1996, Barkana1999, Bullock2000}. However a baryon feedback solution leads to another form of tension, the existence of halos that are seemingly `too-big-to-fail' to produce galaxies \citep{TBTF2011, Griffen2016}.









With the advent of the first high-resolution hydrodynamical simulations in 
which the stellar components of galaxies could be resolved, it became apparent
that realistic galaxy formation required additional heating, such as stellar feedback, to prevent over-cooling \citep{Katz1991,Ceverino2009,Hopkins2012,ReyRaposo2017}. 
The sources of this energy injection, include star formation, supernovae, black holes and feedback from active galactic nuclei (AGN), represent physical processes on scales, often not resolved in cosmological simulations \citep{Somerville2015}. 
Large dynamical scales are required to resolve such feedback processes and large scale structure simultaneously, which significantly stretches computational resources, forcing a compromise between volume and resolution. 
Typically small-scale processes such as star formation, are treated  
with so-called 'sub-grid' methods. Sub-grid methods incorporate simplified models that aim to capture the essential physics of these phenomena while relaxing the need for resolution. 
For example, calculating the details of a single supernovae explosion
is extremely computationally expensive, and is, in itself an outstanding astrophysical problem. Yet supernovae feedback is essential in 
regulating stellar formation in galaxies \citep{Ceverino2009,ReyRaposo2017}, and can possibly induce core formation in dark matter halos \citep{Pontzen2012}.





Feedback mechanisms from baryonic physics are widely proposed to offer solutions to the small scale problems of $\Lambda$CDM, although, unfortunately, 
it has been found that the effect of these sub-grid models of the same 
physical process can vary between different implementations and resolution \citep{niftyI,niftyII,niftyIII,niftyIV,niftyV}. For example, different simulation groups investigating core formation in isolated dwarf halos find contradictory results with each other, some finding cored dwarf halos `all the way down' \citep{Read2016}, no cores at all \citep{Fattahi2016}, or core formation is suppressed at certain scales \citep{Tollet2016Nihao4, Fitts2017}.
However, recent simulations have shown that some of the  problems of $\Lambda$CDM, such as  too-big-to-fail, can be curtailed with more modern codes and state-of-the-art baryon and feedback models \citep{Sawala2016, Wetzel2016}.
 

While there has been significant focus upon feedback from baryonic processes, 
the possibility of processes originating from the `dark sector', such as self-interacting dark matter (SIDM), fuzzy dark matter (FDM), warm dark matter (WDM) and annihilating dark matter, have been neglected in simulations in the past. If present, however, such feedback will have an impact on structure formation and evolution.
For example, self-interacting dark matter \citep{Aarssen2012} has been proposed as a complement to baryon physics in alleviating some of the small scale discrepancies \citep{Spergal2000}. 
In high density environments, a weak self-scattering cross section can cause dark matter to experience pressure like forces, resulting in thermalised cores in the inner regions of halos. At the same time SIDM acts more like collision-less CDM in the low density field, where scattering events are rare, thus replicating the same large scale predictions. 
In recent years a wide range of studies utilising N-body implementations of SIDM have converged upon cross sections $\sigma / m \approx 0.1-10$  $\textrm{cm}^2$ $ \textrm{g}^{-1}$ in dwarf galaxies that are consistent with observations of small structure \citep{Valli2018,Ren2018}, but also leave large scale structure predictions of $\Lambda$CDM intact. Separating core formation by SIDM or baryonic means remains an active area of research \citep{Vogelsberger2014,Fry2015, Dooley2016, Kamada2017, Tulin2018_review, Robertson2018}. Recently \cite{Robles2017} have found that SIDM can produce cores in small dwarfs with stellar masses M$_{\star} < 10^{6} $ M$_{\odot}$. At these scales core formation that would be typically be induced by stellar feedback is cut off by UV photo-ionisation. 

Fuzzy dark matter models, such as ultra-light axions \citep{Hu2000}, have also been proposed as models where their physics can leave marks on the Universes' small scales. New N-body simulations that feature these have recently been produced and their implications are being investigated \citep{Veltmaat2016, Veltmaat2018, Du2018, Nori2018}.

Theoretically a dark matter candidate called Weakly Interacting Massive Particles (WIMPs), self-annihilate in the early Universe and provide an explanation for the dark matter abundance observed today (i.e $\Omega_{m} \approx$ 0.3), assuming an early universe cross section $\langle\sigma v\rangle = 3 \times 10^{-26}$ $\textrm{cm}^3$ $\textrm{s}^{-1}$ \citep{Steigman2012}. Many such particles have been independently hypothesised in particle physics to solve different theoretical problems in the Standard Model, and so are seen as viable candidates for these thermal relics \citep{2010Feng}. As such, significant effort has been directed towards observing the annihilation/decay signals produced from various astrophysical sources containing high dark matter densities such as the Galactic Centre and dwarf spheroidal galaxies. In recent years, a number of studies in conjunction with CMB measurements have constrained the annihilation rate to levels nearing thermal relic cross section, for models where the dark matter particle mass is less than 60 GeV and decay into single standard model channels, ruling out those simple thermal relic WIMPs models \citep{ackermann2014dark,2015Ackermann,Albert2017}. Ground based experiments also place limits at TeV ranges, although not yet near the thermal cross-section \citep{Profumo2016, Oakes2017, Rinchiuso2017}. Relaxing some of the simplifying, albeit restrictive, assumptions about WIMP annihilation results in weaker constraints, for example, multiple decay channels, or the presence of 2 $\rightarrow$ 3 channels, result in softened annihilation spectra which are more difficult to detect. Simply allowing decay into a combination of the well studied single 2$\rightarrow$ 2 channels, weakens the constraints such that we cannot yet rule out WIMPs as light as 2-6 GeV in simple generic thermal dark matter \citep{Leane2018}. Multi-stepped cascade decays within the dark sector can weaken Gamma ray limits by 1-2 orders above the thermal rate for particles as light as $\sim$ 1 GeV, dependent on the number of cascades and standard model final states \citep{Elor2016}, and the presence of 3 $\rightarrow$ 2 `Cannibal dark matter' interactions are relatively unconstrained in the velocity dependant case \citep{Pappadopulo2016}. When cross sections vary with velocity, a generic feature of many models, the cross section limits from the Fermi Gamma-ray telescope are severely weakened as the astrophysical J factors depend on the velocity distribution functions of matter, and weaken cross section constraints by as much as 3 orders of magnitude, above the thermal rate \citep{Zhao2016,Zhao2018}. Other model complications exist as well, but more complications generally only adds more uncertainty to the cross-section constraints. Taken in aggregate this means that the case for GeV WIMP dark matter is yet not as pessimistic as sometimes implied, see \cite{Leane2018} for a discussion on the case for GeV WIMPs. 

Below $\approx$ 1 GeV in the MeV ranges, the current strongest limits on s-wave annihilation arise from the era of recombination, at z $\approx$ 1100, via imprints on CMB \citep{Planck2016}
\begin{equation}
   \frac{f_{eff} \langle \sigma v \rangle}{m_{\chi}} < 4.1 \times 10^{-31}  \;\rm{cm}^3\;  \rm{s}^{-1} \; MeV^{-1},
	\label{eq:plancklimit}
\end{equation}
where $f_{eff}<1$ is a parameter that relates the energy released that can effectively increase the ionisation fraction, thus strongly restricting the simple s-wave thermal WIMP scenario in the sub-GeV range. A number of missions like e-ASTROGAM and AMEGO will be able to probe the the sub-GeV annihilation products directly today with Fermi Gamma-ray telescope like sensitivity \citep{Bartels2017}. Voyager I direct measurements of cosmic ray data however are able to rule out thermal relic WIMPs below about 300 MeV today \citep{Boudaud2017} assuming single decay product annihilations and models of Galactic cosmic ray propagation. 
Sub-GeV dark matter could arise in non-standard thermal production mechanisms or suppression of the cross section at early times \citep{Choquette2016, Xiang2017} or non-WIMP models. Like the higher energy Gamma ray detection experiments, minor extensions to the simple WIMP scenario can drastically alter the constraints, for example p-wave annihilation at the sub-GeV ranges is much less constrained by the CMB and well above the thermal cross sections \citep{Diamanti2014}.
While the WIMP scenario provides a convenient picture of dark matter production from thermal freeze out, it is not the only dark matter model and it is unclear how the previous limits would apply to generic dark matter models. The theoretical uncertainties in the dark sector and the nature of dark matter make it difficult to rule out anything but simple models or specific candidates.


Another less explored avenue of investigation into dark matter annihilation raises the question if any energy released by the decay and subsequent interaction of the annihilation products could play an active role in late structure-formation \citep{Ascasibar2007}. The expected interaction products in the form of particles and photons deposit their energy in to the nearby baryonic material of galaxies and the intergalactic medium, and a number of studies show that Dark Matter Annihilation Feedback (DMAF) could affect gas flows, cosmic rays and early star formation \citep{Ripamonti2007,Natarajan2008,Wechakama2011,Schon2015,Schon2018}. 




To fully understand the impact from DMAF we require numerical simulations to properly capture the non-linear structure growth and to date there has not yet been a self-consistent, hydrodynamic, cosmological simulation including the effects of DMAF from generic dark matter models. In \citet[][hereafter Paper I]{Iwanus2017}  we presented the self-consistent hydrodynamic code that includes the effect of DMAF (which we briefly review in \ref{sec:method}) and tested in the case of isolated toy halo models. In this second paper we present the first cosmological simulations containing DMAF, evolved from perturbations at z $=100$ and report on the differences brought on by DMAF compared to non-cooling simulations focusing on the properties of halos. This paper runs as follows, in section \ref{sec:method} we briefly review our DMAF code implementation and our halo finding code and describe the set up of our simulation boxes. In section \ref{sec:Results} we investigate changes to the simulations in the particle properties, halo catalogues and halo profiles and conclude our study in section \ref{sec:conclusions}.



\section{ Methods }
\label{sec:method}
\subsection{DMAF Implementation}

Full details of the code implementation into \textsf{GADGET-2} are provided in paper I, but here we briefly give an outline. We utilise the standard SPH kernel to estimate the density of dark matter surrounding a gas particle,

\begin{equation}
    \rho_i = \sum_{j=1}^{N}  M_j W(r_{ij},h_i),
	\label{eq:sph_density_estimate}
\end{equation}
where $M_{j}$ is the mass of a nearby dark matter particle, $W$ the smoothing kernel, $r_{ij}$ is the particle separation and $h_{i}$ the smoothing length. The DMAF specific power released by the gas particles in our simulation is 

\begin{equation}
   \frac{du_{i}}{dt} = \frac{\langle\sigma v \rangle c^{2}}{m_{\chi}}\frac{\rho_{\chi}^{2}}{\rho_{g}},
	\label{eq:gas_injection}
\end{equation}
where  $\langle\sigma v \rangle$ is the thermal cross section, $c$ is the speed of light, $m_{\chi}$ is the dark matter particle mass, $\rho_{g}$ is the particles baryonic density and  $\rho_{\chi}$ is the estimated dark matter density as seen locally by the gas particle. To account for dark matter particles annihilating away DM particles can experience mass loss given by 

\begin{equation} \label{eq:masschange}
\begin{split}
\frac{dM_i}{dt} & = -\frac{\langle\sigma v \rangle}{m_{\chi}}\rho_{\chi} M_{i},
\end{split}
\end{equation}
where $M_{i}$ is the mass of the N-body dark matter particle and here $\rho_{\chi}$ is the local dark matter density\footnote{Although we found this effect to be negligible for the cross sections used in this work, in paper I and so was turned off after finding negligible differences in low resolution simulations. It may still be relevant in future work with hidden dark sector models.} \citep[,see Paper I and][for more details]{Monaghan1992}. We assume the mean energy deposition length of the particles is contained within the SPH smoothing length i.e., all energy injection is localised.

We initialise our simulations using \textsf{N-GenIC} \citep{Grossi2009, Springel2015}, though a fully self-consistent simulation should also include the effects of DMAF in the early Universe. However we are primarily focused on the non-linear regime to explore DMAF on late forming structures and so we ignore the effect of DMAF on the initial conditions, but they could result in increased abundances of dwarf halos at mass scales of about $\approx$ M$_{\odot}$ \citep{2006Bertschinger}, well below the resolutions we study here. More complicated models like late-decoupling scenarios with additional SIDM might affect the power spectrum and affect structure at dwarf scales, similar to warm dark matter \citep{Bringmann2016, Binder2016}. It should be noted that in general, the thermally averaged cross section in the early universe is not necessarily the same during late time structure formation. Generic models dark matter models have velocity and spatial dependencies on the particle environments which will differ for late and early annihilation, but in this work we restrict ourselves to a single constant velocity cross section.

\begin{figure*}
	\includegraphics[width=2\columnwidth]{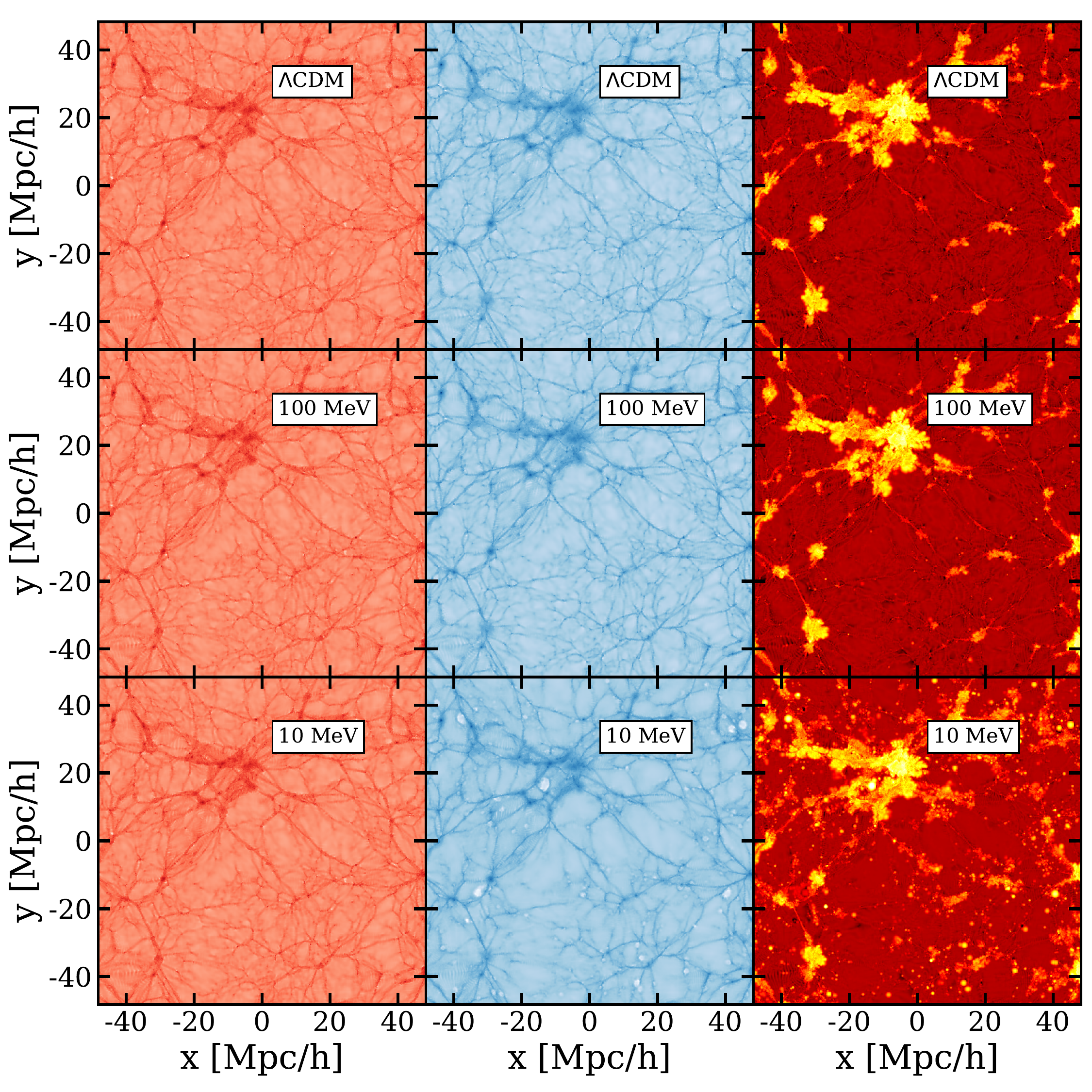}
    \caption{A 2D-Kernel SPH image of some of our simulations. Each row shows a slice of a $\Lambda$CDM control simulation (top row) as well as simulations with an annihilating dark matter model with m$_{\chi}$ = 100 MeV (middle) and an a stronger 10 MeV model (bottom) all with the thermal relic cross section a standard thermal cross section $\langle \sigma v \rangle = 3\times 10^{-26}$. The left column images (red) show the dark matter density, middle (blue) shows the gas density and thermal energy is shown on the right columns (yellow/red). The 100 MeV model has only impacted strongly the small scale structure making differences hard to see, however the 10 MeV model now effects structures large enough to be seen and results in a much smoother field with less filamentary structure.}
    \label{fig:sim_images}
\end{figure*}

\subsection{Halo Catalogues: \textsc{VELOCIraptor} \& \textsc{treefrog}}
\label{sec:halo_finding} 
We generated our halo catalogues using \textsc{VELOCIraptor}, formerly \textsc{STructure-Finder} \citep[][Elahi et al., in prep]{Elahi2011}\footnote{Open source code available at https://github.com/pelahi/VELOCIraptor-STF}. The code identifies candidate halos first with a 3D friends-of-friends (3DFOF) algorithm, and then applies a 6DFOF algorithm to cleanse the catalogue and separate halos connected by spurious particle bridges. Substructures within the field objects are identified by first identifying particles that appear to be dynamically distinct from the background halo particles, i.e. particles which have a local velocity distribution that differs significantly from the mean. These outlier particles are then linked using a phase-space FOF. This approach is capable of not only finding subhaloes, but also tidal streams surrounding subhaloes as well as tidal streams from completely disrupted subhaloes \cite[][]{Elahi2013a}. For this analysis, we are only interested in subhaloes, i.e., self-bound objects, and thus require particles in substructures to have potential energy to kinetic energy ratios of at least $0.95$.
\par 
Linking the structures between different catalogues in time we used the code \textsc{TreeFrog} which is part of the \textsc{VELOCIraptor} package. At its core, \textsc{TreeFrog} matches two objects at different times based on a merit function of the number of shared unique particle IDs between each simulation \cite[see][and Elahi in prep., for more details]{2013SrisawatSUSSING,Elahi2018a}.

\begin{figure*}
\includegraphics[width=2\columnwidth]{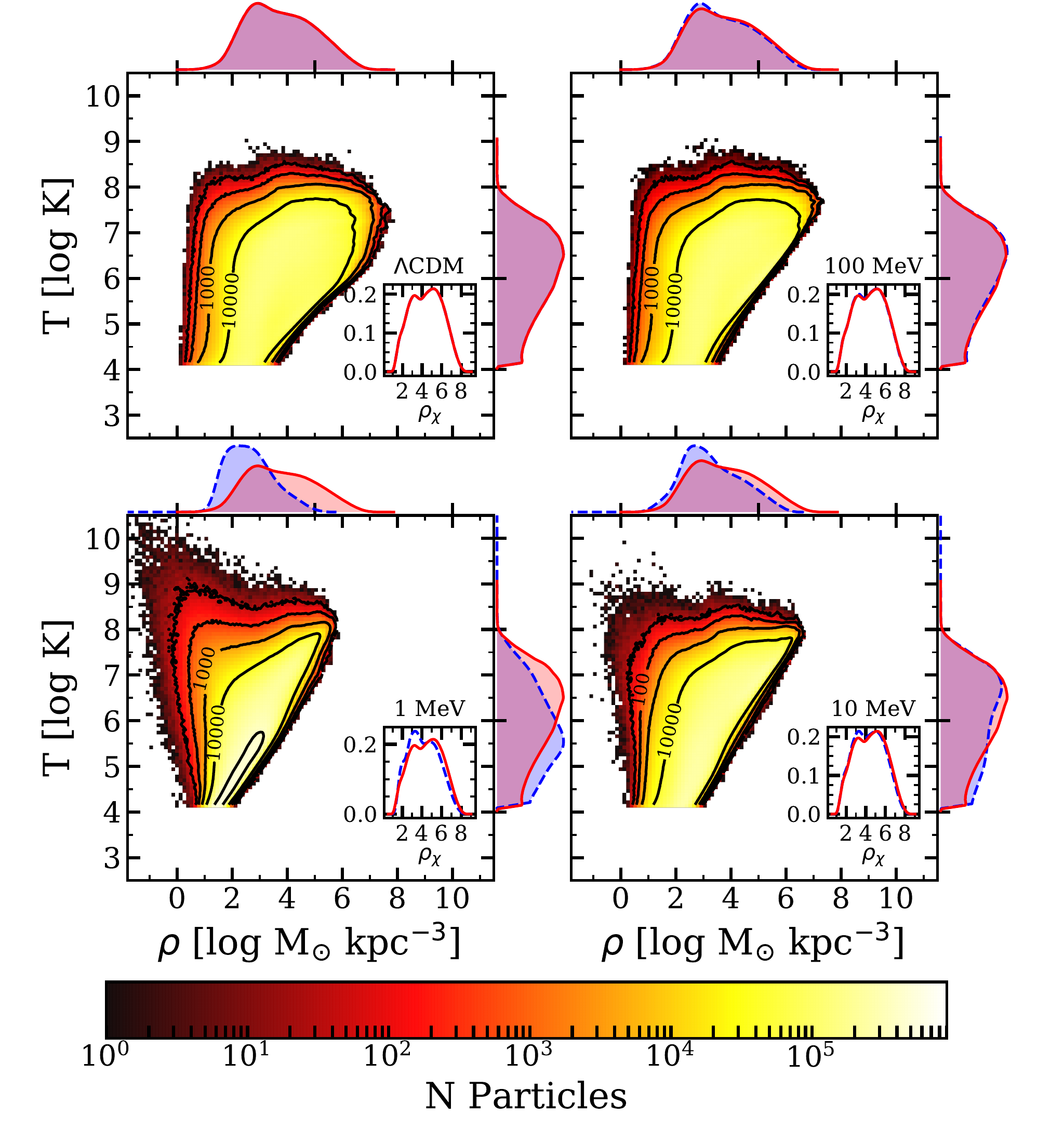}
    \caption{The 2D Histogram showing the temperature and density phase space distribution of the gas in the  $\Lambda$CDM, 100 MeV, 10 MeV and 1 MeV simulations (clockwise). We show the marginal pdfs of the gas density (top add-ons) and the temperature (right add-ons) in in dashed lines compared to the $\Lambda$CDM case (solid line). For clarity in the figures we have shown the distribution for only the particles with T > 10$^4$, our minimum temperature floor. The insets similarly show the corresponding distribution of dark matter particle densities (blue) and comparison with the $\Lambda$CDM DM density distribution (red).}
    \label{fig:rho_u}
\end{figure*}

\subsection{The Simulations}

Our simulations consist of 100 $h^{-1}$ Mpc co-moving length periodic boxes, containing N = 512$^3$ N-body gas and dark matter particles (see Table \ref{tab:Nbody_table}). Our simulation volume was chosen to resolve smaller halos as our previous study and tests in paper 1 indicated that they were more susceptible to the effects of DMAF --- largely due to their decreasing binding energies and higher concentrations. Smaller volumes however reduce the amount of cosmic variance of a simulation, but can also affect the properties of small halo populations due to reduction in tidal interactions by larger clusters. We chose 100 $h^{-1}$ Mpc as a compromise between size and resolution. These parameters at the highest resolutions, allowed us to find bound halos as small as about $10^{10}$ $h^{-1}$ M$_{\odot}$.

The same initial conditions were rerun multiple times under different dark matter models with varying mass and a corresponding control simulation of a pure non-cooling hydrodynamics (no stars, AGN etc.), allowing us to directly track any differences due to DMAF. Our cosmological parameters are taken from the Planck collaboration \citep{Planck2016}. For ease of comparison to the rest of literature, which is dominated by thermal WIMP astrophysical signatures, we have set a constant relic-like cross section and varied the particles mass (see Table \ref{tab:Cosmo_table}). However, with our efficient and localised energy injection scheme, see equation \ref{eq:gas_injection}, decreasing the mass is equivalent to raising the cross section, so the models here should be considered more general than the thermal relic WIMPs. For example, a 100 GeV model can also represent a standard s-wave annihilation with mass of 1 TeV and at a higher cross section of $\langle \sigma v \rangle = 3\times 10^{-25}$ $\rm{cm}^3$ $\rm{s}^{-1}$ which are not yet strongly constrained even for single species, s-wave annihilation. 

We note that although there are strong constraints on light $\sim$ sub MeV dark matter models in the simple WIMP paradigm, it is by no means the only one, and the effects of a generic annihilating dark matter species injecting energy at sites of high $\rho_{\chi}^2$ during structure formation should be thoroughly understood in exploratory N-body simulations before the study of more complex models with p-wave or exotic velocity dependence, multiple-species DM, co-annihilation etc., much like SIDM studies have progressed into studying non-constant cross sections.  In future work, we will generalise our DMAF scheme to study more complex dark matter models with sub-dominant components, non-constant cross sections, late forming dark matter, etc. For this reason we choose to study DMAF for a number of dark matter masses in the standard WIMP thermal freeze-out scenario, including masses ruled out by the CMB \citep{Planck2016}, specifically the 100 keV model. This extreme heating model provides a `stress-test' of the code, allowing us to identify numerical limitations of the scheme as well as aiding in the interpretation of other models. Additionally, such low masses are not definitively ruled out in population suppressed or non-standard production models as in \cite{Heeck2017} and \cite{Berlin2018}.



\begin{table}
	\centering
	\caption{The \textsf{GADGET-2} simulation settings used for our largest resolution simulation.}
	\label{tab:Nbody_table}
	\begin{tabular}{lccr} 
		\hline
		Simulation Settings & Value\\
		\hline
		Box length & 100 $h^{-1}$ Mpc \\
		Particles N & 2$\times$512$^3$\\
        DM N-body mass	& 5.38 $\times$ 10$^8$ $h^{-1}$ M$_{\odot}$  \\
        Gas N-body mass & 0.848 $\times$ 10$^8$ $h^{-1}$  M$_{\odot}$ \\
		$\epsilon_{soft}$ & 9.76 $h^{-1}$ kpc  \\
        N$_{sph}$ & 40 \\
        T$_{min}$ & 10$^{4}$ K \\
        z$_{init}$ & 100 \\
		\hline
	\end{tabular}
\end{table}

 \begin{table}
	\centering
	\caption{Our cosmological parameters are taken from the the Planck 2015 measurements \citep{Planck2016}}
	\label{tab:Cosmo_table}
	\begin{tabular}{lccr} 
		\hline
		Cosmological Parameter & Value\\
		\hline
		$\Omega_{\Lambda}$ & 0.6911 \\
		$\Omega_{m}$ & 0.3075 \\
		$\Omega_{b}$ & 0.0486 \\
        $\sigma_{8}$ & 0.8159 \\
        $h$ & 0.6774 \\
        $n_{s}$ & 0.9667 \\
        $\langle\sigma v\rangle$ & $3\times 10$$^{-26}$ cm$^3$ s$^{-1}$ \\
        $m_{\chi}$ & 100 keV, 1 MeV, 10 MeV,   \\
         & 100 MeV, 1 GeV, 100 GeV   \\
		\hline
	\end{tabular}
\end{table}

\section{Results}\label{sec:Results}
We show the matter distribution in Figure \ref{fig:sim_images}, where the red and blue panels show a slice of the smoothed density fields of dark matter and gas respectively, while the last yellow-red column shows the specific thermal energy. At the scales presented here, there is little perceptible difference between the 100 MeV simulation and the $\Lambda$CDM control simulation, showing that the large scale structure is seemingly unaltered for moderate heating levels, though we will see smaller scale effects later. Ramping up the the annihilation rate by decreasing the dark matter particle mass to m$_{\chi}$ = 10 MeV visibly affects the distribution, smoothing out much of the smaller filamentary structures. For the models with even stronger annihilation, large DM overdensities appear to lack gas. As we further increase the rates of annihilation with 1 MeV and 100 keV, we see large bubbles of hot low density gas begin to form, some of which are also visible to a lesser degree in the $m_{\chi} =$ 10 MeV simulation. These bubbles form in regions where high density dark matter structures inject large amounts of thermal energy into low density gaseous regions. 

These bubbles form even in simulations using smaller and global time stepping schemes. They are not a result of time-step numeric artifacts that can be diminished by using careful time-stepping methods as addressed by \cite{Saitoh2009}. Instead these form due to our assumption that the gas is always efficiently and locally capturing the DMAF energy, even in low density gas. A simple check was performed by repeating lower resolution runs but now adding a simple switch where DMAF heating is turned off in gas particles where densities are lower than the cosmic mean; see Appendix \ref{appendix:bubbles}. This strongly suppressed the growth of these bubbles and the following results that appear in the rest of this paper were found to be insensitive to the time-stepping method as well as the DMAF cutoff switch, thus do not affect the conclusions in the following sections. This is because the bubbles form in areas of low density hot gas, making them gravitationally insignificant and unbound to the halos structures within our simulations; again, see Appendix \ref{appendix:bubbles}. 
\[\]
\subsection{Particle properties}\label{Particle properties}
In Figure \ref{fig:rho_u} we show the temperature and density phase space distribution of the gas particles at z = $0$, where we calculated the temperature of a particle as 
\begin{equation}
	T = \frac{\mu m_p (\gamma - 1)}{k_B} u,
    \label{eq:temp_conv}
\end{equation}
and we assumed the standard \textsf{GADGET-2} settings of mean molecular mass of $\mu$ = 1.22, $\gamma$ = $\frac{5}{3}$, $m_p$ is the mass of a proton, k$_B$ and $u$ is the specific thermal energy of the particle. We find that the particles in high density environments have been preferentially heated due to annihilation feedback, as can be seen by the straightening of the lower right edge of the distribution in the 10 MeV case and a smaller change in the 100 MeV case. Overall however, although there is a slight increase in gas particle numbers at high temperatures above 10$^8$ K, there is a much larger deficiency of particles at temperatures between 10$^6$ and 10$^8$ K and a large increase in the number of colder gas particles below 10$^6$ K. The apparent net effect of annihilation feedback is to prevent condensation and to expel gas particles from high density environments where now fewer but hotter gas particles generate enough pressure to push against the infalling gas which remains cold. In the cases with annihilation stronger than 10 MeV, we see little or no gas exists at high densities as DMAF was strong enough to prevent any gas condensation into the dark matter halos. The insets of each panel in Figure \ref{fig:rho_u} show the corresponding PDFs of the dark matter particle densities. They are only slightly affected in the 100 MeV simulation but show a decrease in the number of dark matter particles at high densities in the 10 and 1 MeV simulation, suggesting that gas changes due to DMAF have gravitationally imprinted onto the dark matter as well.

\begin{figure*}
	\includegraphics[width=2\columnwidth]{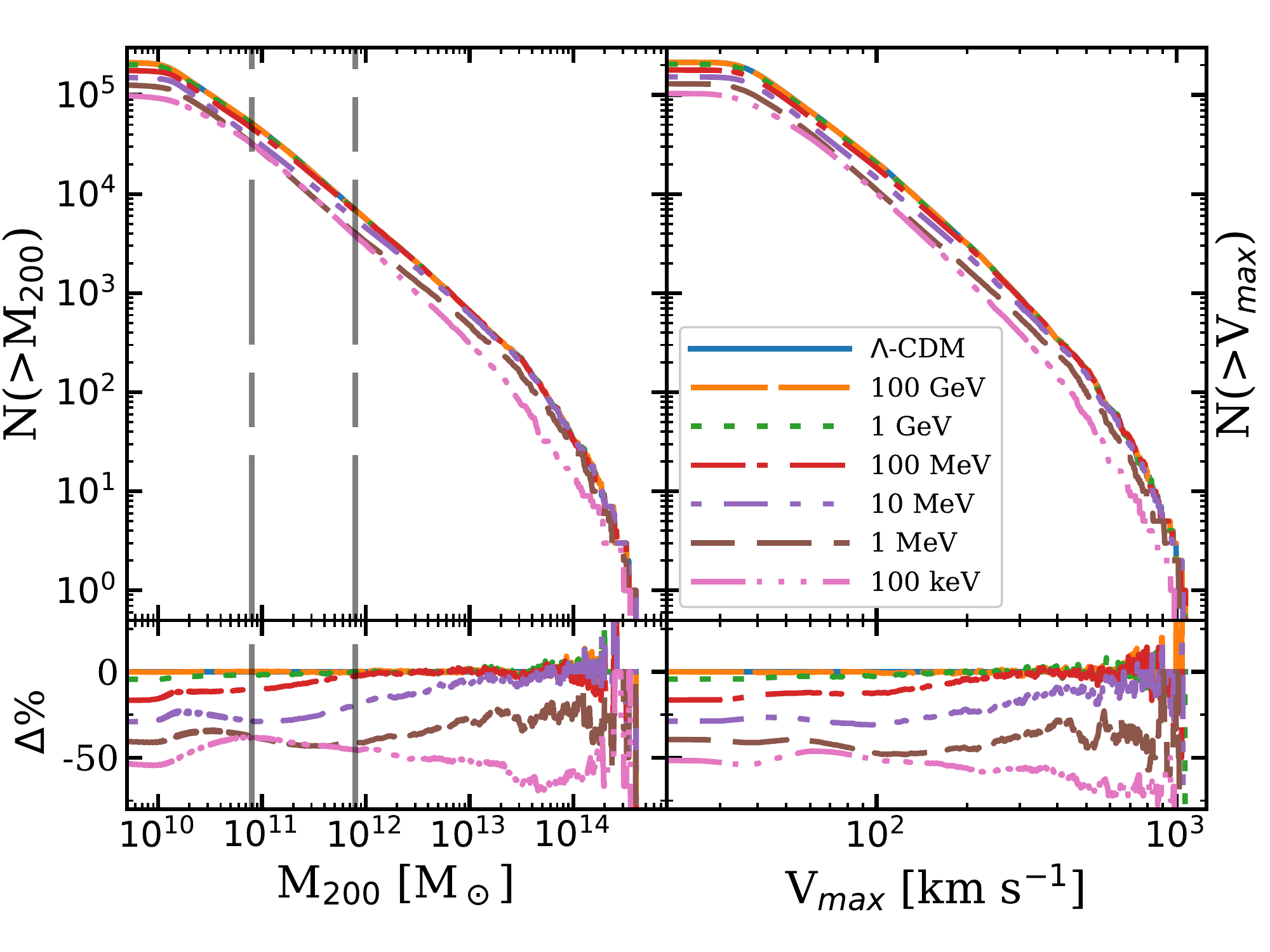}
    \caption{The Halo Mass Function (left) at redshift $z = 0$ and the Halo Velocity Function (right) of our simulations. The residual plots show the percentage difference in the HMF/HVF compared to the control $\Lambda$CDM simulation. For the middle annihilation rate (m$_{\chi}=100$ MeV), DMAF tends to lead to suppression of halo abundance at lower masses, as gas accretion has been prevented in smaller concentrated halos, delaying their growth. Halos with masses > 10$^{13}$ M$_{\odot}$ do not show reduction in abundance except for the strong annihilation models with m$_{\chi}=100$ keV and m$_{\chi}=1$ MeV, where the annihilation is drastic enough as to slow the growth of even the large halos. The m$_{\chi}=100$ MeV model shows a reduction in abundances for halos below 10$^{12}$ M$_{\odot}$, and we see that the m$_{\chi}=1$ GeV model is starting to show sensitivity to DMAF. }
    \label{fig:HMF}
\end{figure*}

\begin{figure*}
 	\includegraphics[width=2\columnwidth]{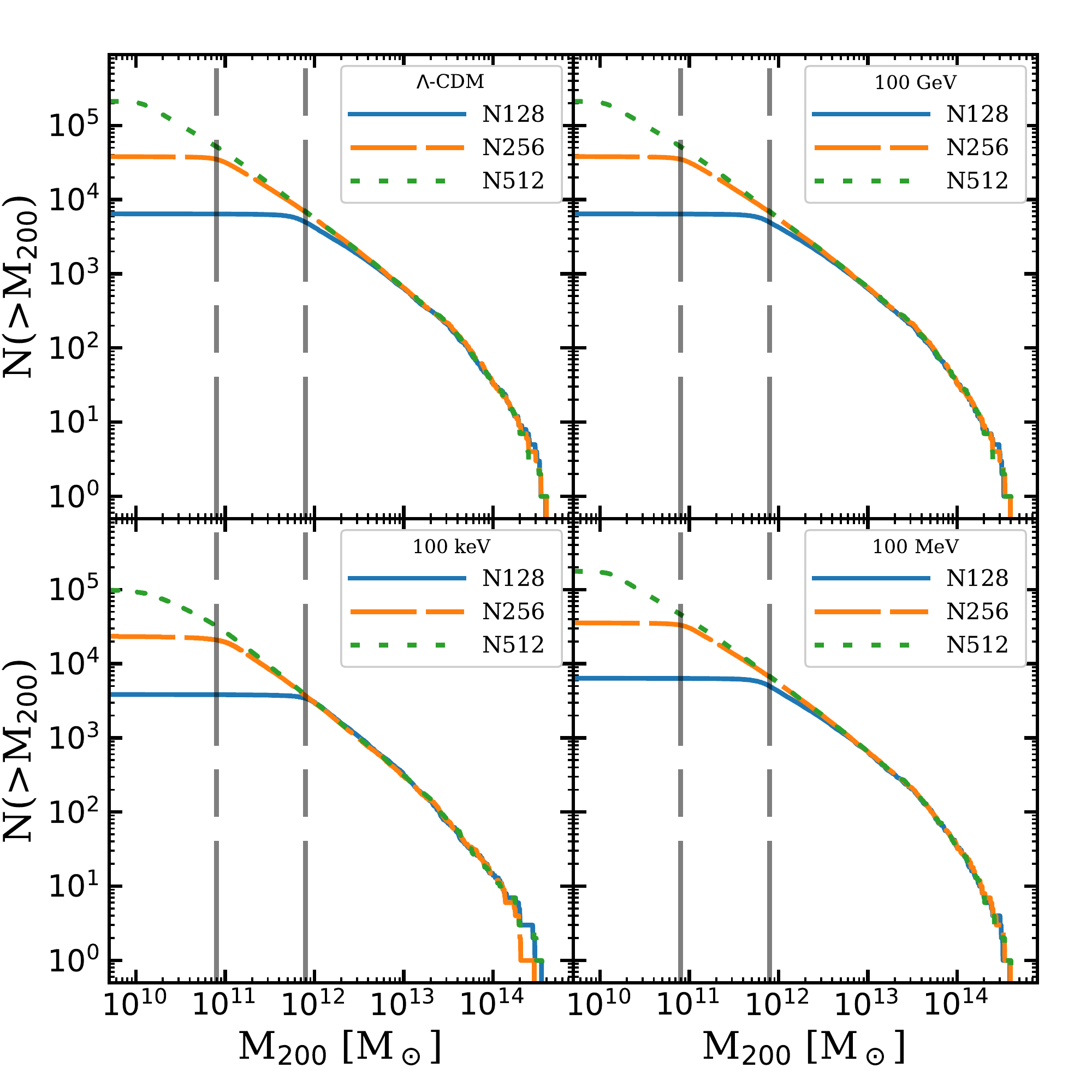}
    \caption{The HMFs for all 4 models sampled at resolutions with N= 2$\times$128$^3$, 2$\times$256$^3$ and 2$\times$512$^3$ particles. Most simulations show good convergence properties in our simulations for all the DMAF models. The grey  dashed lines show the mass equivalent of 100 and 1000 dark matter particles respectively in the 512$^3$ case.} 
    \label{fig:HMF_convergence}
\end{figure*}

\begin{figure*}
	\includegraphics[width=1.77\columnwidth]{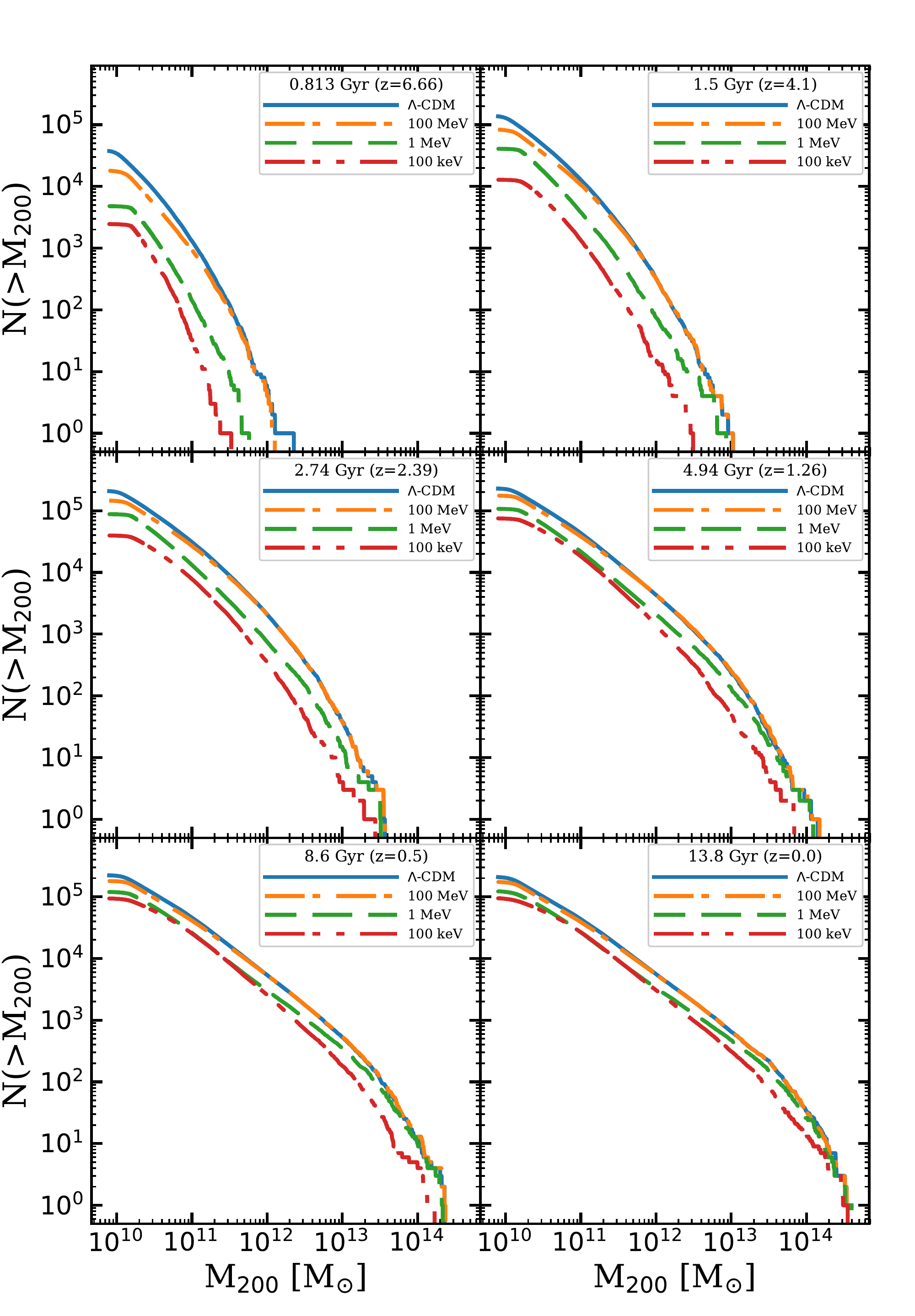}
    \caption{ The Halo Mass Function (HMF) time evolution of our 4 simulations. As seen at $z=6.66$ the differences between the DMAF simulations and the control $\Lambda$CDM simulation are more pronounced at early times due to DMAF slowing/preventing the accretion of gaseous material into the halos. These differences becomes less pronounced (especially at high masses) as we approach $z=0$ where the largest of the halos in all but in the most extreme 100 keV simulation have negligible mass and abundance differences. This 'catch up' can be understood by remembering that large cluster growth becomes dominated by mergers at late times which DMAF appears to do little to prevent.}
    \label{fig:HMF_vs_z}
\end{figure*}

\subsection{Halo Populations}\label{sec:Population}
We quantify the mass of our halos using the common virial radius definition such that,
\begin{equation}
	M_{200} = 200 \rho_{c} \frac{4 \pi}{3} R^{3}_{200},
    \label{eq:virial_mass}
\end{equation}
that is the mass enclosed within a radius, $R_{200}$ that contains an average density 200 times greater than the critical density of the universe, $\rho_{c}$. The maximum circular velocity is calculated as  

\begin{equation}
	V_{\rm{max}}^2 = \rm{max}\Bigg(\frac{GM(r)}{r}\Bigg),
    \label{eq:Vmax}
\end{equation}
where M(r) is the enclosed mass within radius r.

We show the Halo Mass Function (HMF) and Halo Velocity Function (HVF) in Figure \ref{fig:HMF} as well as the associated deviation from the control simulation in the bottom panels. During convergence testing, (see Figure \ref{fig:HMF_convergence}), comparison with lower resolution simulations showed little deviation from their higher resolution counterparts when looking at masses above about 1000 dark matter particles (dashed vertical line), and only minor differences near 100 particles for all simulations. In Paper I and other studies \citep{Ripamonti2007,Schon2015,Schon2018} we had concluded that small concentrated halos tended to be more sensitive to the effects of DMAF due to their high concentrations and lower gravitational binding energy. However the residuals for our strongest feedback model still show large deficits in high mass halo abundance. The 100 GeV model shows little change overall whilst the 1 GeV and 100 MeV models show agreement with our conclusions from Paper I, that the abundance of low mass objects is suppressed\footnote{Although it does begin to rise up again below M$_{200}$ $\approx$ 10$^{11}$ M$_{\odot}$ when the halos only consist of $\approx$ 100 particles for some of the models.}. The onset of this abundance reduction increases with stronger annihilation, occurring just below a `critical halo mass' near 10$^{12}$ M$_{\odot}$ for $m_{\chi}$ = 100 MeV, 10$^{13}$ M$_{\odot}$ for $m_{\chi}$ = 10 MeV and shows reduction at all scales for the strongest models $m_{\chi}$ = 1 MeV and $m_{\chi}$ = 100 keV, the critical mass being above our largest halo mass.   


Figure \ref{fig:HMF_vs_z} sheds some light on the origin of these differences by showing the evolution of the HMF through time. For m$_{\chi}$ = 100 MeV, the differences were larger at high redshift but decreased with time. In the $m_{\chi}$ = 10 MeV case we see that at early redshifts the abundance is reduced at all scales. The injection of energy and suppression of gas infall has increased the formation time of these structures but for large objects (M$_{200}$ > 10$^{13}$ M$_{\odot}$) the delay does not seemingly prevent the largest halos from reaching their full size and we see only small differences in mass abundances by z = 0. For smaller objects however, the DMAF induced delay is more enduring and results in the suppression of low mass objects we saw in Figure \ref{fig:HMF}. In the m$_{\chi}$ = 100 keV model this delay is significant enough to have induced a suppression at all scales, only a few halos managing to catching up. It appears that DMAF in the absence of other feedback has introduced new dynamics into structure formation which can be understood by remembering that the primary way small halos grow at early times is due to the accretion of outlying material, while the formation of large clusters and galaxies is dominated by the mergers and accretion of smaller halos. DMAF slows the accretion of gas, reducing the growth rate of these small early structures but does little once the large structures begin to grow through mergers. In the m$_{\chi}$ = 100 keV and m$_{\chi}$ = 1 MeV simulation we see that DMAF from this extreme model is able to delay the merger growth to such a degree that it does not catch up at the the high mass end.

\begin{figure}
	\includegraphics[width=\columnwidth]{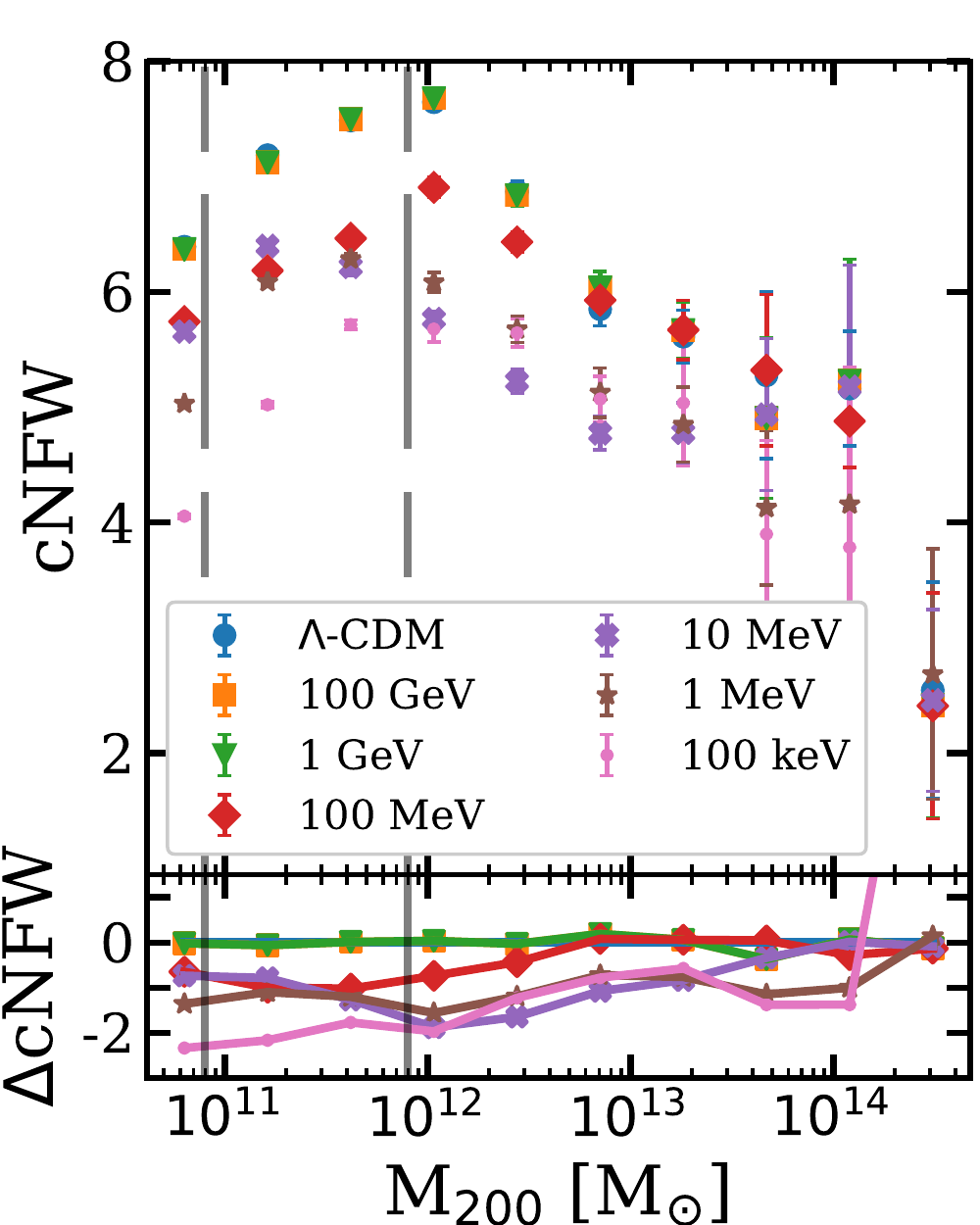}
    \caption{The concentrations of our halos, calculated with equation \ref{eq:prada_c}, under different DMAF models. The error bars are the SEM estimate of each bin and for reference we show two grey dashed lines at a mass equivalent of 100 and 1000 dark matter particle masses and higher masses are better resolved. The lowest panel shows the residual differences of each model compared to the $\Lambda$CDM case.}
    \label{fig:mass_vs_c}
\end{figure}

\subsection{Halo Properties}

The density profiles of halos are reasonably characterised by Navarro-Frenk-White (NFW) halos \citep{Navarro1996}, which follow a profile given by, 
\begin{equation}
    \rho (r)=\frac{\rho_0}{ \left(\frac{r}{r_{s}}\right)\left(1 + \frac{r}{r_{s}} \right)^2},
	\label{eq:nfw}
\end{equation}
where $r_s$ is a scale radius, $\rho_0$ is a density scale. NFW profiles are often parametrised by their mass, see equation \ref{eq:virial_mass} and a concentration parameter given by
\begin{equation}
	c = \frac{R_{200}}{r_{s}}.
    \label{eq:cNFW}
\end{equation}
Concentration parameters in this work are not directly calculated by finding the best fit parameters of the averaged spherical profile. Instead we follow \cite{Prada2012} where we assume an NFW profile, for which the following relation holds.
\begin{equation}
	\frac{V^{2}_{\rm{max}}}{GM_{200}/R_{200}} = \frac{0.216c}{\rm{ln}(1+c) - c/(1+c)}.
    \label{eq:prada_c}
\end{equation}
We plot the $c-M$ relation of our halos in Figure \ref{fig:mass_vs_c}. The scatter in the distribution is quite large and so we show the standard error of the mean to more clearly show the significance of the shift in the distribution centres. In the stronger feedback models, we see that the concentrations are systematically lower at all masses and the effect is an increase in smaller halos, at least up to the dashed vertical line where the halos are sampled with particle numbers less than 1000. These changes can occur through two main modes, DMAF directly altering the equilibrium structure of the halos by the increased thermal energy and lower densities in the gaseous component, but also because of the delayed halo formation time, as the c-M relationship is normally a function of redshift as well \citep{Okoli2017}. Early forming structures tend to have higher concentrations as they are able to effectively accrete the outlying materials at larger radii and so the lower concentrations could be attributed partially attributed to delayed halo formation.

In Paper I (Appendix B for details), we showed the importance of resolving the inner core of halos (i.e below $r_{s}$) in DMAF simulations as the high density cores are expected to be a major source of annihilation. Cusped halos that are considered unresolved below a scale $r_{con}$ will therefore underestimate the total amount of energy injected by the halo by a factor of about $\frac{r_{con}}{r_{s}}$ for an NFW halo, r$_{con}$ can be determined by the `Power criterion' \citep{Power2003}. The lack of a resolved cusp in these poorly sampled halos may explain why the trend for the HMF and concentrations reverses at poorly resolved low mass halos.


We investigated the halos spin parameter as defined by \cite{Bullock2001}

\begin{equation}
	\lambda = \frac{J_{200}}{\sqrt{2}M_{200}V_{200}R_{200}},
    \label{eq:spin_param}
\end{equation}
but we found no clear sign of an increase or decrease brought on by DMAF.

\subsubsection{Gas in Halos}

\begin{figure}
	\includegraphics[width=\columnwidth]{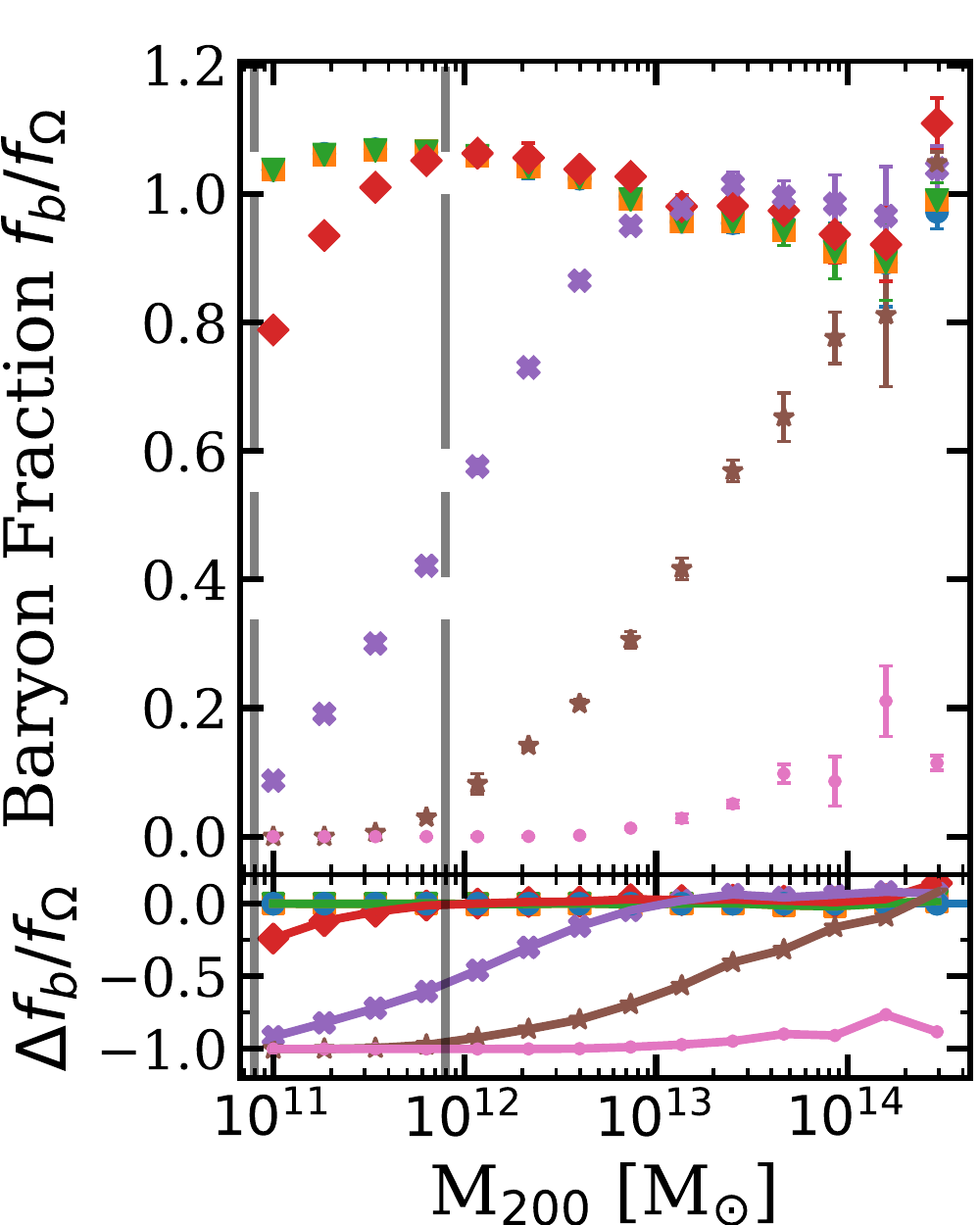}
    \caption{The baryonic fraction, $f_{b}$ relative to the overall cosmic fraction $f_{\Omega} = \Omega_{b}/\Omega_{\chi}$, of the halos at z = 0. The error bars show the standard error of the mean in each bin. The fraction is very sensitive to annihilation feedback most strongly at a critical mass of about 10$^{13}$ $M_{\odot}$ for the 100 MeV simulation, although the largest halos seem to be more gas rich in comparison to the $\Lambda$CDM counterparts, likely because there is now more gaseous material unbound to the smaller objects.}
    \label{fig:baryon_fraction}
\end{figure}

The gas fraction of our halos is defined as 
\begin{equation}
	f_{b} = \frac{m_{g}N_{g}}{m_{g}N_{g}+m_{\chi}N_{\chi}},
    \label{eq:baryon_fraction}
\end{equation}
where $N_{g}$ and $N_{\chi}$ are the number of gas and DM particles bound to the halo weighted by their N-body particle mass, respectively. We find a strong DMAF signature in the baryonic fraction as shown in Figure \ref{fig:baryon_fraction} relative to the the cosmic baryon fraction $f_{\Omega} = \Omega_{b} / \Omega_{\chi}$. The median fraction is heavily reduced in halos that have masses less than a `critical mass' the depends on the strength of the annihilation, above which the baryon fraction returns to near normal. In the 1 MeV model only the largest halos at about $\approx$ 10$^{14}$ M$_{\odot}$ manage to hold onto their gas but drops off to 10$^{13}$ M$_{\odot}$ for the 10 MeV model and just below 10$^{12}$ M$_{\odot}$ for the 100 MeV simulation. We did not see significant gas depletion in the 1 GeV model although given the spacing of the stronger models critical mass it may simply occur just below our simulation resolution. Actually the largest halos seem to be more gas rich in comparison to the $\Lambda$CDM counterparts, this is likely because there is now more gaseous material available for accretion by the large halos as the small ones were unable to capture their own gas.

\begin{figure}
	\includegraphics[width=\columnwidth]{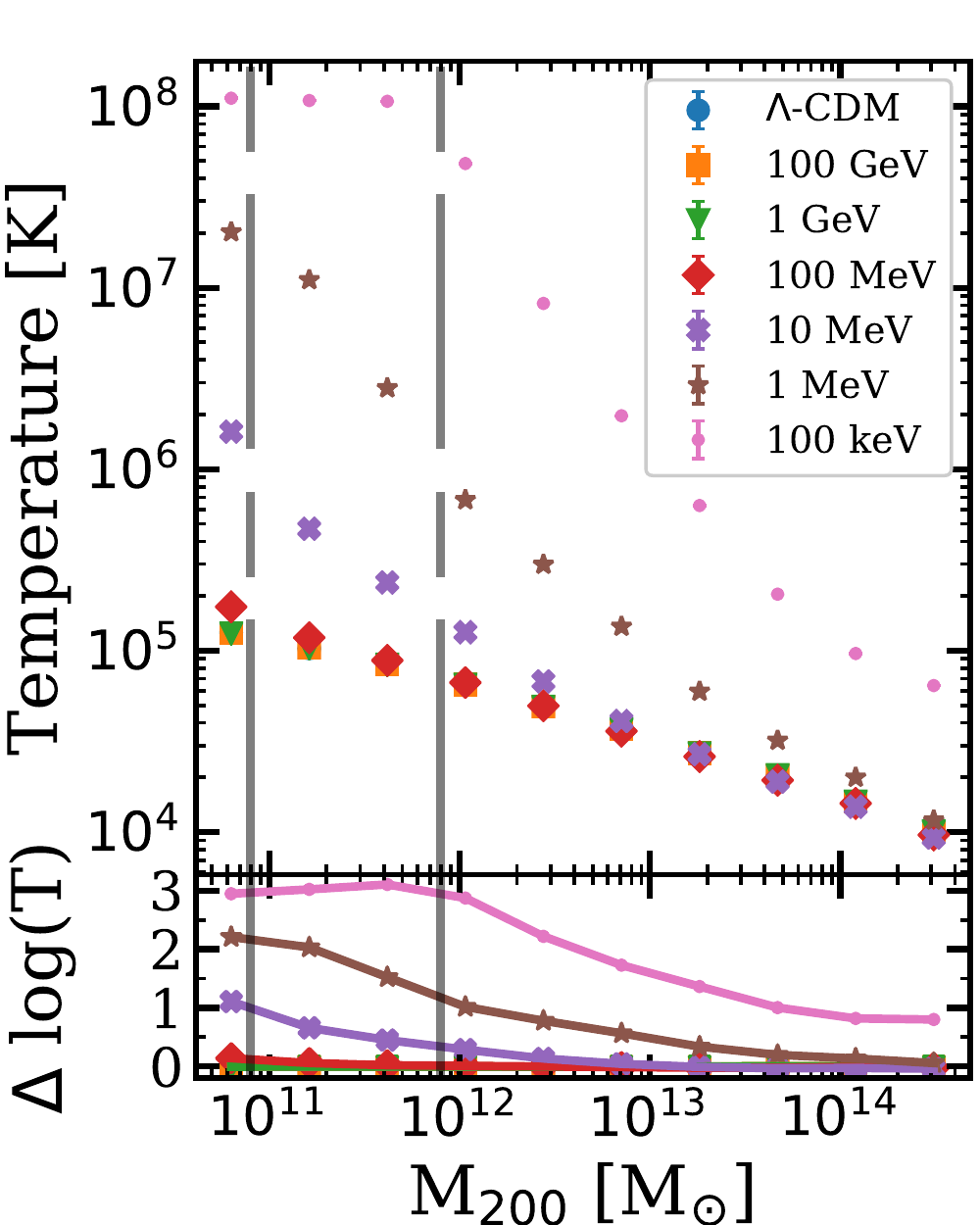}
    \caption{The average temperature of the halos gas as a function of the mass M$_{200}$. We see that the 100 MeV simulation is able to more strongly heat up the gas in less massive halos. The 100 keV simulation has similarly heated up what little gas remains in their halos, see Figure \ref{fig:baryon_fraction}.}
    \label{fig:Temp_vs_mass}
\end{figure}

In Figure \ref{fig:Temp_vs_mass} we calculate the average temperature of the gas in each halo and then show the median of these halos in their respective bins. DMAF has reduced the gas reservoirs in small halos below about 10$^{12}$ M$_{\odot}$ of the 100 MeV simulation, and the 100 keV simulation contains essentially no gas for halos with masses below 10$^{13}$ M$_{\odot}$, but what remaining gas is left exists at much larger temperatures. These results are consistent with the picture drawn in section \ref{Particle properties}, where a less dense but higher temperature gas is able to prevent condensation of further accreting gas into the halos.

In comparison to our non-cooling simulations, such changes are likely to be washed out in real life to some degree by cooling and other feedback processes. However the strongest promise of being able to detect these changes is in regions where baryonic feedback simulations have shown that competing cooling and heating processes are of similar magnitudes. At lower masses, dwarf halos become susceptible to the effects of UV photo-ionisation \citep{Hoeft2006,Fitts2017} and can prevent the condensation of gas into these dwarf galaxies at a UV heating `critical mass' of about 10$^{10}$ M$_{\odot}$ as at these scales the cooling and dominant UV heating are near cancelled out. Our results show that even effective 100 MeV models can disrupt the properties of these halos above this mass scale and we see a faint sign of slightly reduced abundances in the HMF and HVF for the 1 GeV model. Though we just fell short of the required resolution in this study, we see it as a promising area for DMAF signatures as even here we see effects of DMAF at Galactic scales. Below this critical mass scale, we expect UV heating to be the dominant source of energy, however gas is self-shielding \citep{Krumholz2011} which reduces the UV heating efficiency and is taken into account in many modern simulations, whilst an energy source from a dark matter peak could still inject its energy internally bypassing the shielding and producing differences. 

\subsection{Halo Profiles}
\begin{figure*}
	\includegraphics[width=2\columnwidth]{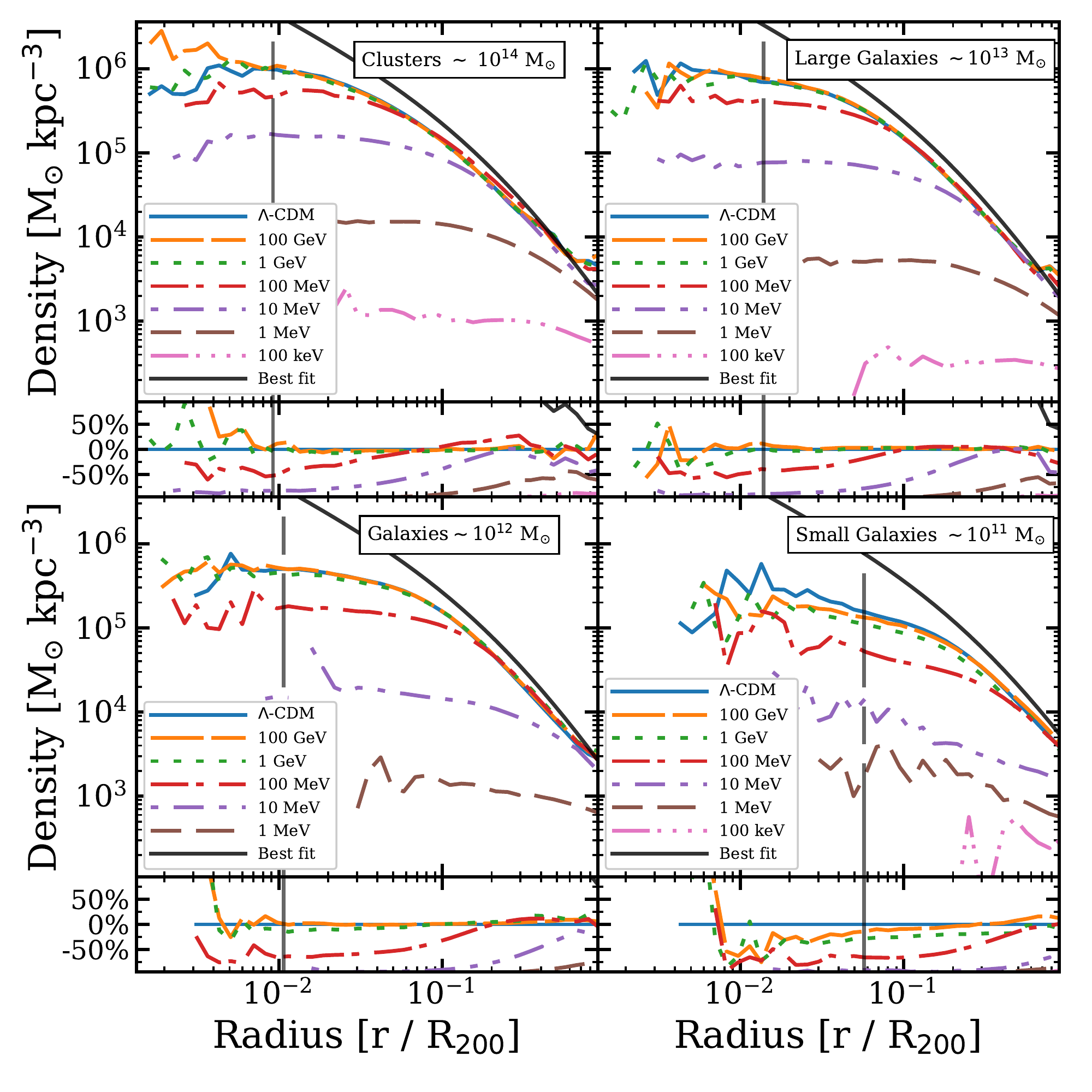}
    \caption{The averaged gas density profiles of 4 halo sub-samples with different masses. The residuals show the percentage density difference compared to the  $\Lambda$CDM simulation. For the largest structures (top panels) the density on the out-skirts shows little difference for 100 MeV dark matter particles but within 0.1 R$_{200}$ from the centre the gas has been severely depleted. The smaller structures (bottom panels) show gas depletion at all radii for the 100 MeV case. At masses of about 10$^{11}$ M$_{\odot}$ our averaged profiles become noisier as it becomes more difficult to achieve accurate centring and density estimates due to a smaller number of particles, but we see that there is a small but systematic deviation even above the typical softening length in the 100 GeV case. The NFW density fit (dark line above) here has been scaled by $\frac{\Omega_b}{\Omega_{\chi}}$.}
    \label{fig:Clusters_Gas}
\end{figure*}

\begin{figure*}
	\includegraphics[width=2\columnwidth]{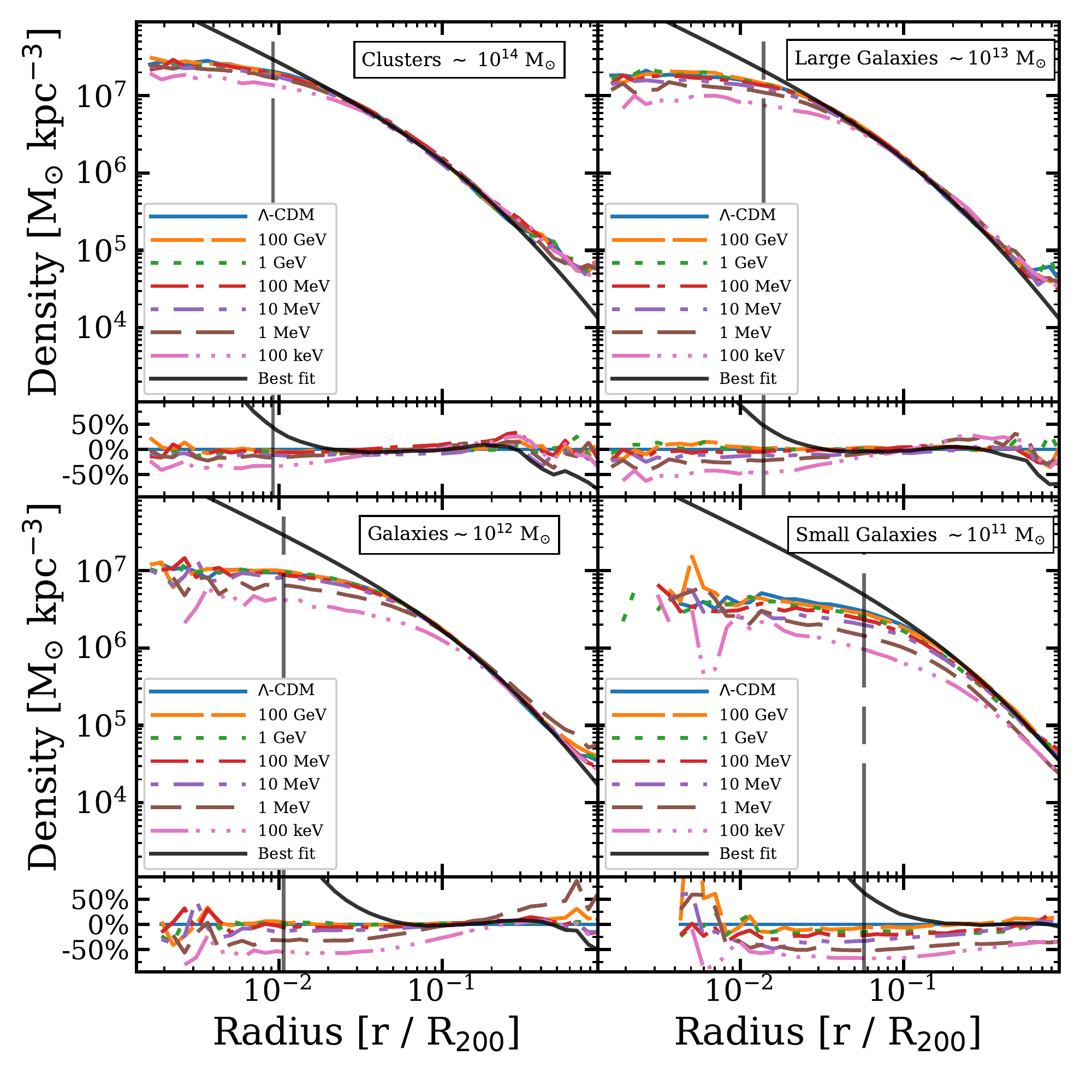}
    \caption{The sample averaged dark matter density profiles of 4 sub-samples of halos with different masses roughly corresponding to cluster, galaxy and dwarf sized halos. The residual panels show the percentage difference compared to the  $\Lambda$CDM simulation (as in Figure \ref{fig:Clusters_Gas}). Also shown is the best fit NFW profile to the $\Lambda$CDM simulation, see equation \ref{eq:nfw}. The greyed dashed line shows the average gravitational softening where the measured densities begin to diverge from the expected profile. }
    \label{fig:Clusters_DM}
\end{figure*}

In Figure \ref{fig:Clusters_Gas}, we show the averaged gaseous density profile as a function of their scaled radius. We show 4 sub-samples of halos with varying M$_{200}$ (sample sizes are $N = 50, 100, 100, 200$) roughly corresponding to our largest clusters, large, medium and small galaxies. The $\Lambda$CDM objects were matched to their DMAF simulation counterparts using \textsc{TreeFrog} and selecting the DMAF simulation halo with the largest merit function,
\begin{equation}
	M_{ab} = \frac{N_{a\cap b}^2}{N_a N_b},
    \label{eq:merit function}
\end{equation}
where $N_a$ and $N_b$ are the number of particles in halos $a$ and $b$ while N$_{a\cap b}$ is the number of uniquely identified particles that appear in both halos. In this work we restricted the merit function to counting only the dark matter particles due to the large depletion of gas in some of the simulations. For reference, we also show the best fit NFW density profile for the dark matter component scaled by $\Omega_{b} / \Omega_{\chi}$ (black profile) in the $\Lambda$CDM simulation, but also show the dark matter density profiles in Figure \ref{fig:Clusters_DM}. The best fit line was calculated by minimising the equation
\begin{equation}
	\chi^2 = \sum_{i=1}^{N} \frac{\rho(r_i) - \rho_{NFW}(r_i ; \rho_{0}, r_s)}{2\sigma_i^2},
    \label{eq:chisq}
\end{equation}
where $r_{_i }$ are the logarithmically spaced radial bins, $\rho_{NFW}$ is given by equation \ref{eq:nfw} and $\rho(r_i)$ and $\sigma_i$ are the average and standard deviation of the density. The grey dashed vertical line shows average R$_{200}$ scaled softening length $\epsilon_{soft}$ of the halos where we expect artificial gravitational softened cores to exist. In our fitting procedure we used 40 logarithmically spaced bins between 10$\epsilon_{soft}$ < r < R$_{200}$ so as not to include the artificial, gravitationally softened cores in the fit. A more careful analysis would also fit for the halo centres as a free parameter but here we simply centred the halos on the minimum potential.

The feedback has significantly altered the structure within the inner radius, (< 0.1 R$_{200}$) of all samples with $m_{\chi}$ >1 GeV DMAF and we see evidence that the density profiles are more sensitive as the average mass of the halo sample is decreased.  In the 10$^{12}$ M$_{\odot}$ sample, (bottom left panel), the 100 MeV model shows a larger gas depletion of about 50 percent up to around 0.1 R$_{200}$. We also now see a small systematic depletion in the 1 GeV model at this mass. For the smallest mass 10$^{11}$ M$_{\odot}$ sub-sample we clearly see the 100 MeV model show stronger gas depletion now reaching out to the edge of the halos, as well as a larger systematic depletion in the 1 GeV and even perhaps the 100 GeV profile of about 10-20 percent for radii greater than the softening. However the halos here are more poorly sampled due to small particle counts which affects the reliability of our centering, R$_{200}$ estimates and density. Nonetheless this depletion appears in both the 1 GeV and 100 GeV model which were persistent even when a slightly different mass range was chosen. There is evidence that the dark matter profile has responded to changes in the gas density as seen in Figure \ref{fig:Clusters_DM}. For the large cluster sized halos there is little change up to the softening lengths, but we see greater depletions of dark matter when the looking at smaller halos with stronger DMAF. This shows that weak DMAF could still alter the halo morphology for small halos without significantly altering the global statistics within a mass bin. For example, the 100 MeV model shows clear depletion of gas in the inner regions  of large galaxy sized objects (10$^{13}$ M$_{\odot}$) but global statistics such as the HMF and gas fraction are not significantly affected at this range.

\section{Conclusions}\label{sec:conclusions}

In this paper we have presented cosmological hydrodynamical simulations that contain Dark Matter Annihilation Feedback, in which sources of energy at places of high dark matter concentrations couple thermal energy into the nearby gas via interaction with their decay products. We have produced a set of 100 $h^{-1}$ Mpc wide box simulations with N=2$\times$512$^{3}$ gas and dark matter particles from identical initial conditions and we examined the distribution of gas and dark matter particles as well as the properties of dark matter halos to explore differences generated by these DMAF events.


Our results from these simulations suggest two main conclusions and ways forwards with simulation studies of the DMAF. We find that small dwarf galaxies and sub-halos are sensitive to DMAF and would show direct evidence of feedback effects owing to their lower gravitational binding energies. In the 100 MeV model for example, the abundance of halos is reduced by about 10 percent at redshift $z = 0 $ for halos with masses less than 10$^{12}$ M$_{\odot}$. The evolution of the Halo Mass Function reveals that the mass scale below which this suppression occurs decreases with time, essentially allowing the HMF at high mass scales to progressively `catch up' with the $\Lambda$CDM halo abundance albeit at delayed time. In our simulations, the 100 MeV case was able to catch up at late times ($z = 0$) for masses above $10^{12}$ M$_{\odot}$, but the delayed formation time due to the strong annihilation in the 100 keV simulation results in almost no halos at any mass being able to catch up to their $\Lambda$CDM abundances at the end of the simulation, leaving about 50 percent less structure at all scales. Examination of our halos properties shows that while DMAF leads to higher temperature gas especially in lower mass halos, the dominant effect is to prevent gas condensation into the halos onto smaller dwarf halos and delaying the growth of the larger structures. The density profiles of our halos show that we expect weaker annihilation models to be able to alter the the morphology of the halos at smaller mass scales, where we see weak signs that even a 1 GeV model can induce small changes at scales of $10^{11}$-10$^{12}$ M$_{\odot}$ in the inner regions of small halos.

As discussed in the introduction, models with dark matter particle masses less than 1 GeV are met with tension from a number of astrophysical tests such as the CMB and Gamma Ray studies. This means the energy injection rates these models provide would only be relevant for exotic models with complicated decay spectra or behaviour that are somehow able to evade said constraints. For the models $\approx$ 1 GeV or greater, this tension with observations is slackened such that they do not rule out even the simple S-wave annihilation WIMPs and our models at these scales show promise in imprinting signatures onto structures at dwarf halo mass scales below $\approx$ 10$^{11}$ M$_{\odot}$.

Although the signs of DMAF are eventually washed out at high mass scales and late times -- therefore harder to detect, the delay in the formation of these galactic structures could result in a difference of their star formation histories and ages thus allowing the possibility of detecting this type of dark sector physics even in the large structures.

\section*{Acknowledgements}
The authors would like to acknowledge the University of Sydney HPC service for providing high-performance computational resources from \textsf{ARTEMIS} which has contributed to the research results reported within this paper as well as the assistance of resources and services from the National Computational Infrastructure (NCI), which is supported by the Australian Government. Nikolas Iwanus is supported by an Australian Postgraduate Award (APA). Florian List is supported by the University of Sydney International Scholarship (USydIS). We would like to thank the anonymous referee for their valued feedback which lead to improvements in the quality of this paper.





\bibliographystyle{mnras}
\bibliography{bib} 



\appendix

\section{Origin of the hot bubbles}\label{appendix:bubbles}

\begin{figure}
	\includegraphics[width=\columnwidth]{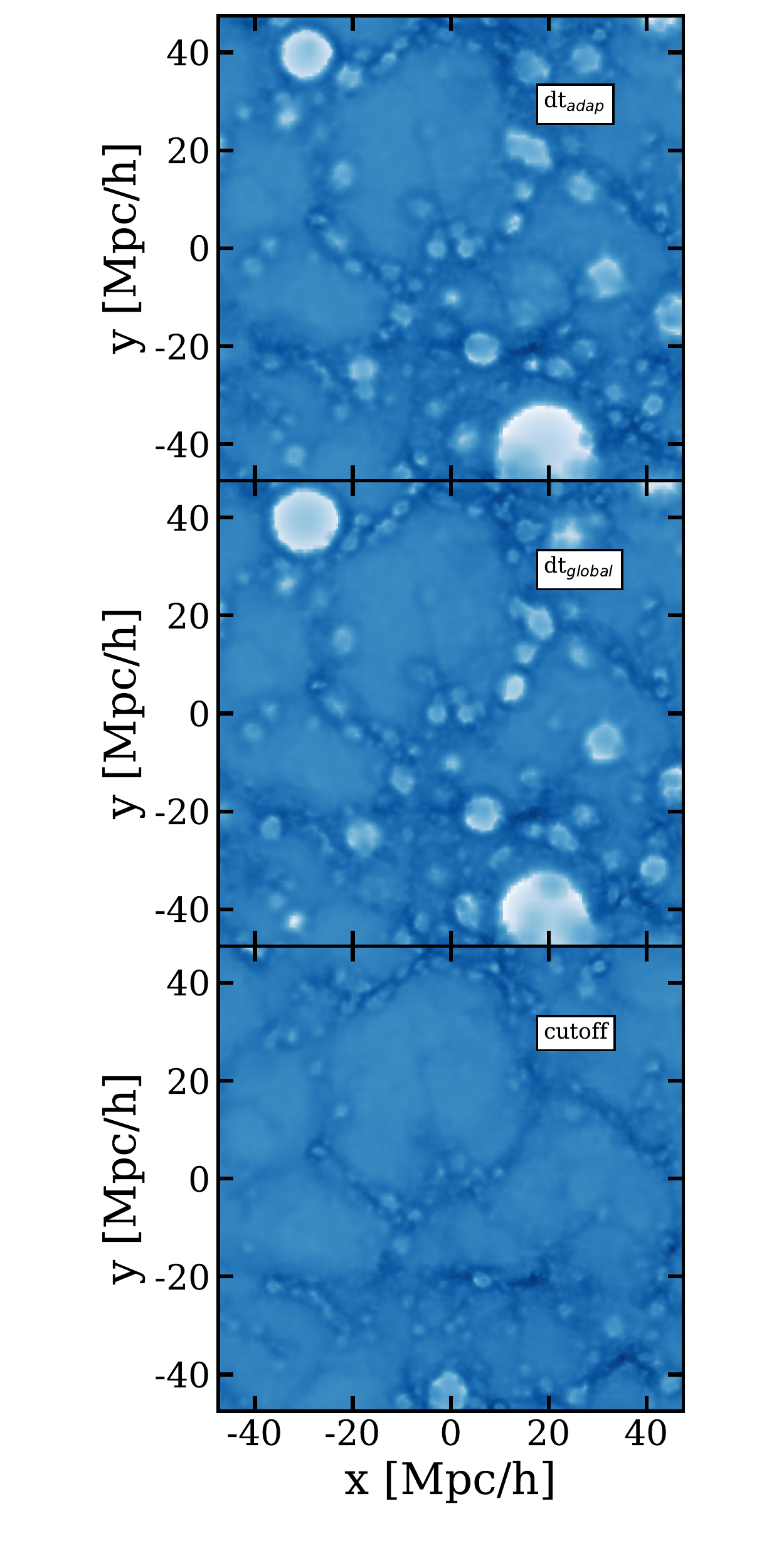}
    \caption{The gas density of three 100 keV simulations with different time stepping schemes. They contain N = 2$\times$128$^3$ particles with the regular \textsf{GADGET-2} adaptive time-stepping scheme (dt$_{adap}$),  global equal time steps (dt$_{equal}$). In addition to a simulation run with a heating cutoff at low densities ($\frac{\rho_{cut}}{\rho{crit}} $ = $0.8  \Omega_{b}$) to suppress the growth and formation of bubbles that form in areas where the gas density is reduced strongly.}
    \label{fig:cutoff_image}
\end{figure}

\begin{figure}
	\includegraphics[width=0.99\columnwidth]{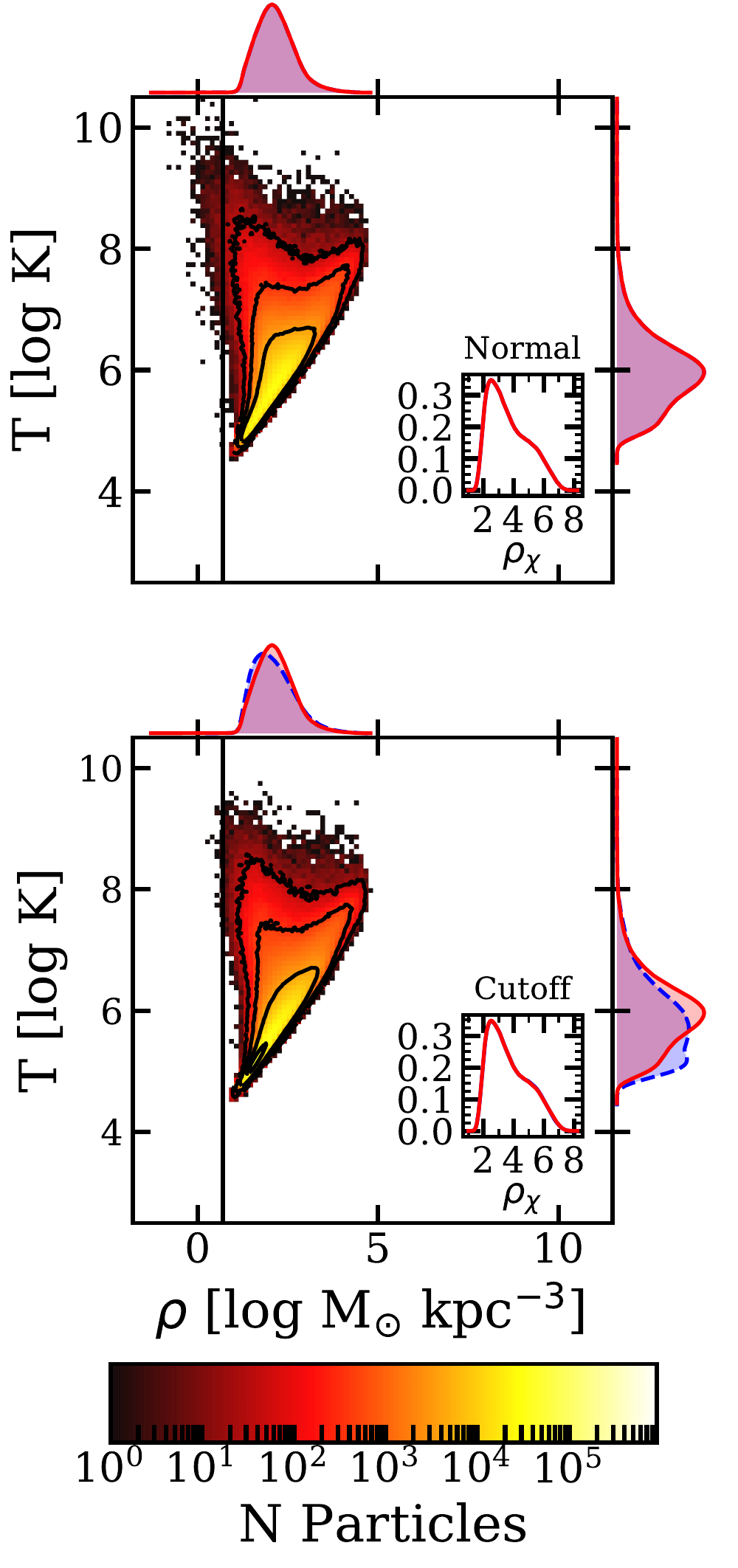}
    \caption{Comparison of two 100 keV simulations, with N = 2$\times$128$^3$ particles with the normal \textsf{GADGET-2} adaptive time-stepping scheme as well as a simulation run with equal time-steps in addition to a heating cutoff switch at low densities ($\frac{\rho_{cut}}{\rho{crit}} $ = $0.8\Omega_{b}$) shown as black vertical line, to suppress the formation of bubbles. The 2D distribution largely retains the same overall shape although we have reduced the amount of hot low density gas particles below the  density cutoff.}
    \label{fig:cutoff_change}
\end{figure}

In simulations with high injection rates, we see the formation of hot bubbles; a 100 keV simulation, see first panel Figure \ref{fig:cutoff_image}, shows the formation of these bubbles at all resolutions while the 10 MeV simulation only formed these bubbles in the  $2\times$512$^{3}$ simulation. The bubbles appear even with stricter time-stepping and global time steps, see the middle panel Figure \ref{fig:cutoff_image}, suggesting they are not simply a result of insufficient time-step limiters such in \cite{Saitoh2009} or \cite{Durier2012}, rather they are due to our assumption that the coupling to the gas is always 100 percent efficient and local. These bubbles originate from areas of the simulation where the gas is depleted relative to the dark matter density (like in the smaller halos). In these areas, the gas receives the same amount of energy but now shared shared amongst less dense gas leading to further rarefaction of a hotter gas which further expands the bubble, this trend continues as the heating is never shut off. 

We have performed a simple check to show that this is the case by implementing a gas efficiency step function. If a gas particle reaches a density lower than the cosmic average, e.g. $\alpha_{cut}$ $\Omega_{b} > \frac{\rho_{i}}{\rho_{crit}}$ then we consider the gas too low density to absorb the products efficiently and it receives no further energy. We have run this on smaller sized simulations with a resolution of 2 $\times$ 128$^3$ particles; see Figure \ref{fig:cutoff_image} bottom panel for the simulation run with $\alpha_{cut} = 0.8$. Although these low density regions still form we found that this density cut off heavily suppresses the formation and growth of these bubbles. We retrieve the same results as expected from lower resolutions of a largely homogeneous gas with no or few gas particles bound to structure. The density phase space distribution is only strongly affects high temperature low-density particles; see Figure \ref{fig:cutoff_change}. We have also checked that the other results in this work, such as the halo mass functions, baryon fractions and temperatures and found sub-percent differences despite the stricter time stepping and percent level differences with the cutoff, smaller then the error of the binning \footnote{We tried a range of cutoffs between 0.3-0.8. Lower cutoffs led to less suppression of the bubbles, but smaller differences.}. 

These bubbles only occur when the gas is sufficiently depleted in the dark matter structure anyway. For example the baryon fraction of our halos is unchanged when the bubbles are suppressed because \textsc{VELOCIraptor} considers these gas particles to unbound to the halos anyway. The halo mass functions are also unaffected because none of the rarefied gas is considered bound and so has little gravitational influence on the structures, other than not being accreted in the first place. For this reason we believe that bubbles we see in Figure \ref{fig:cutoff_change} have minimal impact, the results presented in the rest of the paper are sound; the sharp density contrasts only exist in an extremely rarefied and unbound gas. In addition, for testing purposes we have also implemented DMAF into the more modern code \textsf{GIZMO} which contains time-stepping limiters as in \cite{Durier2012} as well as the pressure formulation of SPH, meshless finite-volume and mass hydrodynamic implementations, see \cite{Hopkins2015,Hopkins2017} and find that the bubbles arise in all configurations, although this \textsf{GIZMO} implementation is less thoroughly tested than our \textsf{GADGET-2} implementation. Investigations are currently underway systematically studying the detailed effects due to different hydrodynamic implementations with DMAF and are anticipated in the next paper of this series.

The bubbles may be diminished with a distributed energy injection scheme and non-localised heating. As the gas density falls off we expect the injected energy to be more spread around out rather than concentrating in these hot, low density bubbles and we aim to study the result of energy injection dispersed over a mean free path in future work. Our simulations are also non-cooling and so would also be aided by thermal conduction, radiative cooling etc. which are likely to help prevent such an extreme build up. These changes are anticipated in future codes but for now these bubbles are only prominent in the strong injection models anyway and do not significantly affect our results within the halos themselves.



\bsp	
\label{lastpage}
\end{document}